\UseRawInputEncoding
\documentclass[superscriptaddress, twocolumn, amsmath, amssymb, aps,pra,letterdraft, notitlepage,longbibliography]{revtex4-1} 
\usepackage{graphicx,graphics,epsfig,subfigure,times,bm,bbm,amssymb,amsmath,amsfonts,amsthm,mathrsfs,MnSymbol}
\usepackage[matrix,frame,arrow]{xypic}
\usepackage[normalem]{ulem}
\usepackage{slashed}
\usepackage{dcolumn}
\usepackage{tabularx}
\usepackage{amsopn}
\usepackage{color}
\usepackage[usenames,dvipsnames,svgnames,table]{xcolor}
\usepackage[english]{babel}
\usepackage{verbatim}
\definecolor{darkblue}{rgb}{0.0,0.0,0.3}
\usepackage[colorlinks=true,
            linkcolor=red,
            urlcolor= darkblue,
            citecolor=blue]{hyperref}

\newcommand{\bea}{\begin{eqnarray}}
\newcommand{\eea}{\end{eqnarray}}

\begin{document}
\title{Initial-state-dependent quantum speed limit for dissipative state preparation: Framework and optimization}

\author{Junjie Liu}
\email{jjliu.fd@gmail.com}
\affiliation{Department of Physics, International Center of Quantum and Molecular Structures, Shanghai University, Shanghai, 200444, China}
\author{Hanlin Nie}
\email{e0547610@u.nus.edu}
\affiliation{Centre for Quantum Technologies, National University of Singapore, Block S15, 3 Science Drive 2, 117543, Singapore}

\date{\today}

\begin{abstract}
Dissipation has traditionally been considered a hindrance to quantum information processing, but recent studies have shown that it can be harnessed to generate desired quantum states. To be useful for practical applications, the ability to speed up the dissipative evolution is crucial. In this study, we focus on a Markovian dissipative state preparation scheme where the prepared state is one of the energy eigenstates. We derive an initial-state-dependent quantum speed limit (QSL) that offers a more refined measure of the actual evolution time compared to the commonly used initial-state-independent relaxation time. This allows for a passive optimization of dissipative evolution across different initial states. By minimizing the dissipated heat during the preparation process, conditioned on the minimization of evolution time using the QSL, we find that the preferred initial state has a specific permutation of diagonal elements with respect to an ordered energy eigenbasis of increasing eigenvalues. In this configuration, the population on the prepared state is the largest, and the remaining diagonal elements are sorted in an order resembling that of a passive state in the same ordered energy eigenbasis. We demonstrate the effectiveness of our strategy in a dissipative Rydberg atom system for preparing the Bell state. Our work provides new insights into the optimization of dissipative state preparation processes and could have significant implications for practical quantum technologies.   
\end{abstract}

\maketitle

\section{Introduction}
The ability to prepare specific quantum states is a fundamental requirement in many areas of quantum information, such as quantum error correction, fault tolerant quantum computation \cite{Nielsen.02.NULL,Bravyi.05.PRA,Fowler.12.PRA,Reed.12.N,Nigg.14.S,Kelly.15.N,Takeda.22.N}, and quantum simulation \cite{Bloch.08.RMP,Lewenstein.07.AP,Haffner.08.PR,Blatt.12.NP,Xiang.13.RMP,Monroe.13.S,Devoret.13.S}. In these applications, a constant supply of ancillary qubits in known low-entropy states is necessary to achieve reliable and efficient processing. Furthermore, quantum computation models that achieve exponential speed-up over classical ones require initialization of quantum systems into pure states \cite{Knill.98.PRL}. In the case of one-way quantum computing \cite{Raussendorf.01.PRL,Walther.05.N}, for example, one should start with a multipartite system prepared in a cluster state. 

Conventional methods for quantum state preparation involve the use of quantum gates that operate in a unitary manner \cite{Long.01.PRA,Plesch.11.PRA,Sun.21.A,ZhangX.21.PRR,ZhangX.22.PRL}. While this approach is compatible with current unitary quantum computation and simulation designs, its implementation is hindered by the accumulation of quantum unitary operation errors during the process \cite{Long.01.PRA,Plesch.11.PRA,Sun.21.A}, as well as the cost of an exponential number of quantum gates or ancillary qubits \cite{ZhangX.21.PRR,ZhangX.22.PRL}. Dissipative state preparation (DSP) \cite{Harrington.22.NRP} represents a departure from traditional state preparation methods. Instead of viewing dissipation as a source of noise to be minimized, DSP exploits it as a resource. By harnessing the dissipation from surrounding environments, DSP enables the generation of desired quantum states as steady states of engineered dissipative evolution processes. Advances in theoretical \cite{Kraus.08.PRA,Diehl.08.NP,Verstraete.09.NP,Leghtas.13.PRA,Kastoryano.11.PRL,Kastoryano.13.PRL,Reiter.13.PRA,Schuetz.13.PRL,Cormick.13.NJP,Mirrahimi.14.NJP,Reiter.16.PRL,Popkov.16.PRA,Puri.17.NPJQI,Tuorila.17.NPJQI,LiD.18.PRA,Zippilli.21.PRL,Zapletal.22.PRXQ} and experimental \cite{Barreiro.11.N,Krauter.11.PRL,Lin.13.N,Shankar.13.N,Kienzler.S.15,Leghtas.15.S,Morigi.15.PRL,Schwartz.16.PRL,Shankar.16.PRX,Lehnert.21.NP,Cole.22.PRL} research have demonstrated the utility of DSP, making it a promising scheme for preparing quantum states.  In addition to the growing body of research supporting the utility of DSP, empirical evidence indicates that this approach may provide significant advantages over other state preparation methods in certain physical systems. For instance, the fidelity scaling achievable using DSP in the preparation of a maximally entangled state of two atoms in an optical cavity surpasses that of any coherent unitary protocol currently available \cite{Kastoryano.11.PRL}. 

While DSP has shown great potential for practical state preparation, it is essential to consider the speed of convergence for a given implementation. 
State-of-the-art DSP builds upon Markovian evolution dynamics described by the Lindblad master equation \cite{Kraus.08.PRA,Diehl.08.NP,Verstraete.09.NP}, but its slow speed due to small damping rates is a significant limitation \cite{Cormick.13.NJP,Breuer.02.NULL}. Faster evolution processes are needed to prevent the accumulation of implementation errors, as well as to suppress heat dissipation and entropy production  \cite{Auffeves.22.PRXQ}. Several active control protocols \cite{LiJ.16.PRA,ChenY.17.PRA,ZhengR.21.PRA,Alexander.20.PRA,WangY.21.QIP,Bao.22.A} have been developed to steer the evolution through well-designed control fields. However, their experimental implementation remains a significant challenge to date. Recently, a passive speed-up strategy utilizing the Mpemba effect \cite{Carollo.21.PRL,Kochsiek.22.PRA} has been proposed as an alternative. This approach avoids using complex control fields and instead tunes initial states such that the slowest decaying mode cannot be excited at the outset.

In the pursuit of minimizing the preparation time of DSP, it is crucial to have a suitable measure of the evolution time of the underlying Markovian process for benchmarking. In many existing studies \cite{Kraus.08.PRA, Verstraete.09.NP, Reiter.13.PRA, Carollo.21.PRL, Kochsiek.22.PRA, Zapletal.22.PRXQ}, the relaxation time, which is defined as the inverse of the Liouvillian spectral gap, is often used as the measure. However, despite its simplicity and rigor, the relaxation time defines a worst-case scenario and does not offer any physical intuition on how to speed up the evolution process. Additionally, it should be noted that the actual evolution time varies between different initial states \cite{WuS.15.JPA, Carollo.21.PRL, Zhou.21.PRR, Kochsiek.22.PRA}. Since the relaxation time is solely determined by the spectrum of the Liouvillian super-operator, it fails to capture the initial-state dependence of the actual evolution time. Given these limitations, a more refined measure for the evolution time is required for DSP.

In this regard, it is worth noting that for unitary state preparation, a quantum speed limit (QSL) has been derived \cite{Bukov.19.PRX} that provides a lower bound for the preparation time and can be used to develop optimal driving protocols. However, a corresponding QSL for DSP has not yet been established. In this work, we derive an initial-state dependent QSL for measuring the preparation time of DSP by utilizing the relative purity \cite{Mendonca.08.PRA, Del.13.PRL, ZhangY.14.SR, Meng.15.SR, Ektesabi.17.PRA, Wu.18.PRA, Liang.19.RPP, Pires.21.PRA} as an information-theoretic quantifier for distinguishing two quantum states. Importantly, the relative purity also serves as an experimentally friendly measure \cite{Giordani.20.NJP}. Notably, the obtained QSL depends solely on the overlap between the initial and final prepared states, enabling a passive optimization over initial states. Our optimization scheme thus distinguishes itself from existing active ones \cite{LiJ.16.PRA, ChenY.17.PRA, ZhengR.21.PRA, Alexander.20.PRA, WangY.21.QIP, Bao.22.A} in that it requires no active controls during the evolution. In addition, it differs from existing passive ones \cite{Carollo.21.PRL, Kochsiek.22.PRA} as it does not require knowledge of the eigen-spectrum of the Liouvillian super-operator. 

To obtain a more comprehensive understanding, we explore the minimization of the quantum speed limit (QSL) while simultaneously accounting for the dissipated heat, which represents the system's energy change during the preparation process \cite{Saito.22.PRL}. This dissipated heat is a critical quantity from an energetic standpoint \cite{Auffeves.22.PRXQ}. Recognizing that the final prepared state is one of the system's energy eigenstates \cite{Kraus.08.PRA} and that only the diagonal elements of the initial states in the energy basis are relevant when evaluating both the QSL and dissipated heat, our investigation concentrates on diagonal initial states in the energy basis for performing optimization. By minimizing dissipated heat while optimizing the QSL, we identify the optimal permutation of a given set of diagonal elements with respect to an ordered energy basis of increasing eigenvalues. The preferred initial state with the optimal permutation exhibits the largest population on the prepared state. Furthermore, the remaining diagonal elements are sorted in an order that is reminiscent of a passive state \cite{Allahverdyan.04.EPL, Uzdin.21.PRXQ} in relation to the same ordered energy basis [cf. Eq. (\ref{eq:opt_rho})].

We illustrate the effectiveness of our optimization framework by applying it to an open Rydberg atom system used to prepare the Bell state \cite{LiD.18.PRA}. Interestingly, the example shows that in the DSP scheme the prepared state can be an excited energy eigenstate, in direct contrast to usual bit reset and information erasure studies (see, e.g., Refs. \cite{Mohammady.16.NJP, Klaers.19.PRL, Miller.20.PRL, Zhen.21.PRL, Saito.22.PRL}) where the final ground state is typically assumed. As a result, the dissipated heat can become negative in DSP with the preferred initial state, indicating that the non-thermal environment injects energy into the system to assist DSP, and moreover, the conventional Landauer principle \cite{Landi.21.RMP} is not applicable to such DSP processes since the system entropy change is instead positive.

The paper is organized as follows. In Sec. \ref{sec:2}, we briefly review the underlying idea of DSP for completeness and state the questions of interest. We then introduce our initial-state dependent QSL based on the relative purity in Sec. \ref{sec:3} and provide optimization details in Sec. \ref{sec:4}. In Sec. \ref{sec:5}, we present an example and demonstrate the effectiveness of our framework. Finally, we conclude our study in Sec. \ref{sec:6}. Two appendices provide complementary derivation details.

\section{Background and question}\label{sec:2}
State-of-the-art DSP harnesses Markovian dissipative evolution processes to prepare useful quantum states \cite{Kraus.08.PRA,Diehl.08.NP,Verstraete.09.NP}. Specifically, DSP utilizes the Lindblad master equation governing the evolution of a reduced system density matrix $\rho_t$
\begin{equation}\label{eq:lindblad}
\dot{\rho}_t~=~-i[H_S,\rho_t]+\sum_{\mu=1}\gamma_{\mu}\mathcal{D}[L_{\mu}]\rho_t.
\end{equation}
Here, $\dot{A}\equiv\frac{d}{dt}A$ is used throughout the study, $H_S$ is the system Hamiltonian, $\gamma_{\mu}\geqslant 0$ is the damping coefficient of channel $\mu$, $\mathcal{D}[L_{\mu}]\rho=L_{\mu}\rho L_{\mu}^{\dagger}-\frac{1}{2}\{L_{\mu}^{\dagger}L_{\mu},\rho\}$ is the Lindblad superoperator with $L_{\mu}$ the Lindblad jump operator for channel $\mu$ and $\{A,B\}=AB+BA$ the anti-commutator between $A$ and $B$. The final stationary state would be a pure one
$|\Phi\rangle$ when the following conditions are fulfilled \cite{Kraus.08.PRA,Buca.12.NJP}
\bea\label{eq:conditions}
&& H_S|\Phi\rangle~=~E_{n^{\ast}}|\Phi\rangle,\nonumber\\
&&L_{\mu}|\Phi\rangle~=~0,~\forall\mu.
\eea
With the above conditions, one can easily check that the density matrix $\rho_f=|\Phi\rangle\langle \Phi|$ is indeed the fixed point of the dynamics with $\frac{d}{dt}\rho_f=0$. We denote $|\Phi\rangle=|E_{n^{\ast}}\rangle$ for later convenience. We remark that the Lindblad operators involved here clearly do not fulfill the detailed balance condition \cite{Horowitz.13.NJP} and $|E_{n^{\ast}}\rangle$ can be an excited energy eigenstate of the system.

Considering a generic initial state in an ordered energy eigenbasis $\{|E_n\rangle\}$ of $H_S$ with eigenvalues sorted in an increasing order, $E_1\leqslant E_2\leqslant\cdots \leqslant E_N$, 
\begin{equation}
\rho_0~=~\sum_{nm}\rho_0^{nm}|E_n\rangle\langle E_m|.
\end{equation}
For our purpose, we decompose it as $\rho_0=\rho_0^d+\rho_0^c$ with diagonal part $\rho_0^d$ and coherence part $\rho_0^c$ respectively reading
\begin{equation}
\left\{\begin{array}{c}
\rho_0^d ~=~\sum\limits_n\lambda_n|E_n\rangle\langle E_n|,\\
\rho_0^c ~=~\sum\limits_{n\neq m}\rho_0^{nm}|E_n\rangle\langle E_m|.
\end{array}\right.
\end{equation}
Here, we denoted $\lambda_n\equiv \rho_0^{nn}$. Noting that in the Lindblad master equation the evolution of populations and coherences in the energy eigenbasis are decoupled \cite{Breuer.02.NULL,Lidar.19.A}, then for a generic initial state, the DSP consists of two independent evolution processes
\begin{equation}
\left\{\begin{array}{c}
\rho_0^d~\Rightarrow~\rho_f,\\
\rho_0^c~\Rightarrow~0.
\end{array}\right.
\end{equation}
Here we are interested in purely the state preparation process and hence limit ourselves to initial states that are diagonal in the energy eigenbasis, namely, $\rho_0=\rho_0^d$ hereafter. 

Evolving the system state to a final pure state also results in changes in the averaged system energy in terms of the dissipated heat $Q$ \cite{Saito.22.PRL},
\begin{equation}\label{eq:heat}
Q ~=~ -\int_0^{T_f}\mathrm{Tr}[H_S\dot{\rho}_t]dt~=~\mathrm{Tr}[H_S\rho_0]-E_{n^{\ast}}.
\end{equation}
Here, we have denoted $T_f$ as the final time with $\rho_{T_f}=\rho_f$. We note that the change in system von Neumann entropy $\Delta S=-\mathrm{Tr}[\rho_0\ln\rho_0]+\mathrm{Tr}[\rho_f\ln\rho_f]=-\mathrm{Tr}[\rho_0\ln\rho_0]=-\sum_n\lambda_n\ln\lambda_n\ge 0$ is fixed once initial populations $\{\lambda_n\}$ are fixed. Taking into account that the dissipated heat depends solely on the diagonal elements of $\rho_0$, we now pose the following intriguing question:
{\it For a given set of populations $\{\lambda_n\}$, where $\lambda_n\geqslant 0$, $\lambda_n\neq\lambda_{m\neq n}$, and $\sum_n\lambda_n=1$, what is the optimal permutation of $\{\lambda_n\}$ with respect to the ordered basis $\{|E_n\rangle\}$ such that the resulting initial state $\rho_0=\sum\limits_n\lambda_n|E_n\rangle\langle E_n|$ can minimize the evolution time of the process $\rho_0\Rightarrow\rho_f$ while simultaneously maintaining the dissipated heat at the lowest possible level?}

\section{Quantum speed limit}\label{sec:3}
To quantify the evolution time from an initial state $\rho_0$ to the final pure state $\rho_f$, we resort to the QSL which provides a lower bound on the evolution time. By noting that $\rho_f$ is pure, we adopt the Uhlmann fidelity, or equivalently in this scenario, the relative purity $\mathrm{Tr}[\rho_t\rho_f]$ \cite{Mendonca.08.PRA,Del.13.PRL,ZhangY.14.SR,Meng.15.SR,Ektesabi.17.PRA,Wu.18.PRA,Liang.19.RPP,Pires.21.PRA} to measure the distance between $\rho_f$ and $\rho_t$, 
\begin{equation}\label{eq:measure}
\mathcal{F}(\rho_t,\rho_f) ~=~ [\mathrm{Tr}(\sqrt{\rho_t}\rho_f\sqrt{\rho_t})^{1/2}]^2~=~\mathrm{Tr}[\rho_t\rho_f].
\end{equation}
We highlight that here we take the final pure state as the reference, in sharp contrast to existing studies \cite{Del.13.PRL,ZhangY.14.SR,Meng.15.SR,Wu.18.PRA,Kobayashi.20.PRA} on QSL which instead chose the initial state as the reference. 

We parameterize the fidelity with an angle $\Theta_t$ changing from a given initial value $\Theta_0\in\Big(0,\frac{\pi}{2}\Big]$ to a final value $\Theta_f=0$
\begin{equation}
\mathcal{F}(\rho_t,\rho_f)~=~\cos\Theta_t.
\end{equation}
We then consider the equation of motion for $\Theta_t$,
\bea
\dot{\Theta}_t &=& \frac{-1}{\sqrt{1-\mathrm{Tr}(\rho_t\rho_f)^2}}\mathrm{Tr}\left(\frac{d\rho_t}{dt}\rho_f\right)\nonumber\\
&=&\frac{1}{\sin\Theta_t}\mathrm{Tr}\left[\left(i[\rho_f,H_S]-\sum_{\mu}\gamma_{\mu}\mathcal{D}^{\dagger}[L_{\mu}]\rho_f\right)\rho_t\right]\nonumber\\
&=&\frac{-1}{\sin\Theta_t}\mathrm{Tr}\left[\left(\sum_{\mu}\gamma_{\mu}L_{\mu}^{\dagger}\rho_fL_{\mu}\right)\rho_t\right].
\eea
In arriving at the second line, we have defined $\mathcal{D}^{\dagger}[L_{\mu}]\rho= L_{\mu}^{\dagger}\rho L_{\mu}-\frac{1}{2}\{L_{\mu}^{\dagger}L_{\mu},\rho\}$. In getting the last line, we have used the properties of $|\Phi\rangle$ such that $[\rho_f,H_S]=0$ and $L_{\mu}\rho_f=0=\rho_fL_{\mu}^{\dagger}$ ($L_{\mu}^{\dagger}\rho_f\neq0$ generally). 

To proceed, we note that the above equation can be rewritten as 
\begin{equation}
\dot{\Theta}_t~=~\frac{-1}{\sin\Theta_t}\mathrm{Tr}\left[\left(\sum_{\mu}\gamma_{\mu}L_{\mu}^{\dagger}\rho_fL_{\mu}\right)(\rho_t-\rho_f)\right]
\end{equation}
by noting $\mathrm{Tr}[L_{\mu}^{\dagger}\rho_f L_{\mu}\rho_f]=0$. Using the Cauchy-Schwarz inequality for operators $|\mathrm{Tr}(A^{\dagger}B)|\leqslant ||A||_F||B||_F$ with $||A||_F=\sqrt{\mathrm{Tr}(A^{\dagger}A)}$ the Frobenius or Hilbert-Schimidt norm, we find from the above equation that
\bea
\left|\dot{\Theta}_t\right| &\leqslant& \frac{1}{\sin\Theta_t}\left|\left|\sum_{\mu}\gamma_{\mu}L_{\mu}^{\dagger}\rho_fL_{\mu}\right|\right|_F||\rho_f-\rho_t||_F\nonumber\\
&\leqslant& \frac{\sqrt{2}}{\sin\Theta_t}\left|\left|\sum_{\mu}\gamma_{\mu}L_{\mu}^{\dagger}\rho_fL_{\mu}\right|\right|_F\sqrt{1-\cos\Theta_t}.
\eea
Here, we have noted that $||\rho_f-\rho_t||_F=\sqrt{\mathrm{Tr}(\rho_t^2-2\rho_t\rho_f+\rho_f^2)}\leqslant\sqrt{2-2\mathrm{Tr}(\rho_t\rho_f)}=\sqrt{2-2\cos\Theta_t}$ due to the properties $\mathrm{Tr}(\rho_t^2)\leqslant 1$ and $\mathrm{Tr}[\rho_f^2]=1$. Integrating both sides of the above inequality from $0$ to the final time $T_f$, we find
\begin{equation}
\left|\int_{\Theta_0}^{\Theta_f}d\Theta_t\frac{\sin\Theta_t}{\sqrt{1-\cos\Theta_t}}\right|~\leqslant~\sqrt{2}\mathcal{A}T_f,
\end{equation}
yielding the following initial-state dependent QSL for $T_f$
\begin{equation}\label{eq:QSL}
T_f~\geqslant~T_{\mathrm{QSL}}~\equiv~\frac{\sqrt{2-2\cos\Theta_0}}{\mathcal{A}}.
\end{equation}
Here, we have defined a non-negative coefficient $\mathcal{A}\equiv||\sum_{\mu}\gamma_{\mu}L_{\mu}^{\dagger}\rho_fL_{\mu}||_F$ which is independent of the initial state and remains fixed once $L_{\mu}$ and $|\Phi\rangle$ are specified. We note that by following the procedures in Ref. \cite{Del.13.PRL} one can obtain a slightly different form of QSL which is, however, looser than $T_{\mathrm{QSL}}$ defined above; see appendix \ref{a:1} for derivation details. It is worth mentioning that (i) $T_{\mathrm{QSL}}$ is independent of dynamics such that one can evaluate it without solving the evolution equation Eq. (\ref{eq:lindblad}), and (ii) Only the numerator of $T_{\mathrm{QSL}}$ depends on the initial state which brings simplicity when performing optimizations. We further note that $T_{\mathrm{QSL}}$ is a monotonic decreasing function of $0\leqslant \cos\Theta_0<1$ and attains its maximum when $\cos\Theta=0$, that is, when $\rho_0$ is orthogonal to $\rho_f$.

\section{Optimization over initial states}\label{sec:4}
We now turn to the question mentioned at the end of Sec. \ref{sec:2}. For a given set of populations $\{\lambda_n\}$ with $\lambda_n\ge0$, $\lambda_n\neq\lambda_{m\neq n}$ and $\sum_n\lambda_n=1$, one can generate a set of diagonal initial states $\rho_0$ each characterized by a specific permutation of $\{\lambda_n\}$ with respect to the ordered energy basis $\{|E_n\rangle\}$. We identify the optimal permutation among them by optimizing two incompatible objectives: the evolution time characterized by the QSL $T_{\mathrm{QSL}}$ [cf. Eq. (\ref{eq:QSL})] and the dissipated heat $Q$ [cf. Eq. (\ref{eq:heat})]; In Appendix \ref{a:2}, we illustrate this incompatibility by showing that minimizing $Q$ alone yields a passive initial state \cite{Allahverdyan.04.EPL,Uzdin.21.PRXQ} which in general does not minimize $T_{\mathrm{QSL}}$ simultaneously as we will see.

We note that minimizing $T_{\mathrm{QSL}}$ amounts to maximizing the relative purity $\mathrm{Tr}[\rho_0\rho_f]$ between initial and final states, hence our optimization is of a Pareto-type \cite{Saito.22.PRL} and one can assign a multi-objective functional $\mathcal{W}=gQ-(1-g)\mathcal{F}(\rho_0,\rho_f)$ with $g\in[0,1)$ a weighting factor. In the present study, we are particularly interested in the case with $g>0$ and $(1-g)/g\gg 1$ such that while both the heat and relative purity are optimized the latter takes precedence over the former. In general, solving the optimization problem with respect to the functional $\mathcal{W}$ subject to an arbitrary initial state should resort to numerical treatments. However, with a given set of populations $\{\lambda_n\}$, we find that $\mathcal{W}=g\sum_n\lambda_nE_n-gE_{n^{\ast}}-(1-g)\lambda_{n^{\ast}}$ is a linear function of $\{\lambda_n\}$; $n^{\ast}$ marks the index corresponding to the final prepared state (see Eq. (\ref{eq:conditions})). Hence, we can tackle the optimization as follows: We first maximize the relative purity by selecting the population among $\{\lambda_n\}$ on the state $|E_{n^{\ast}}\rangle$. We next determine the arrangement of the remaining populations on the remaining energy eigenstates by minimizing the dissipated heat.

%
\begin{figure*}[thb!]
 \centering
\includegraphics[width=2\columnwidth]{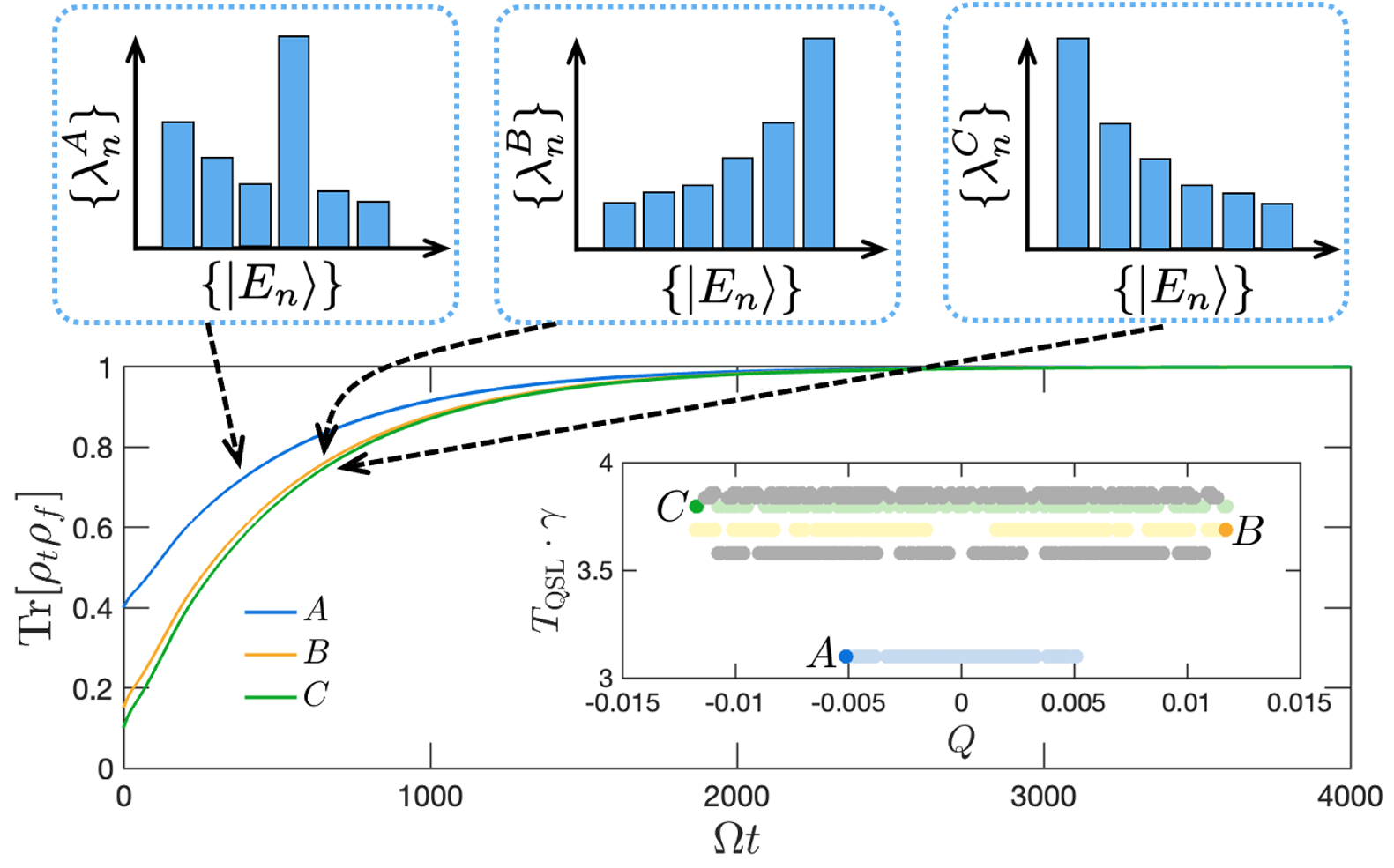} 
\caption{Relative purity $\mathrm{Tr}[\rho_t\rho_f]$ as a function of evolution time obtained by solving Eq. (\ref{eq:lindblad_R}) with three initial conditions $\rho_0^{A,B,C}=\sum_n\lambda_n^{A,B,C}|E_n\rangle\langle E_n|$. We indicate the arrangements of permutations $\{\lambda_n^{A,B,C}\}$ (see detailed definitions in Eq. (\ref{eq:abc})) with respect to the ordered energy basis $\{|E_n\rangle\}$ on the top. Permutation $A$ is the optimal one according to Eq. (\ref{eq:opt_rho}). Inset: we show the corresponding values of $T_{\mathrm{QSL}}$ [cf. Eq. (\ref{eq:QSL})] and $Q$ [cf. Eq. (\ref{eq:heat})] for {\it all} possible permutations of the same set of populations as in Eq. (\ref{eq:abc})), with permutations $A,~B$ and $C$ highlighted. In the calculations, we set $\Omega=2\pi$ MHz as the unit.
}
\protect\label{fig:2}
\end{figure*}
To maximize the relative purity, we consider the following function 
\begin{equation}
\mathrm{Tr}(U\rho_0U^{\dagger}\rho_f)
\end{equation}
with $U$ a permutation operation. Noting that both $\rho_{0,f}$ are diagonal in the energy eigenbasis and $\rho_f$ is pure, we find from the general results in \cite{Stoustrup.95.PRL} that 
\begin{equation}
\mathop{\mathrm{max}}[\mathrm{Tr}(U\rho_0U^{\dagger}\rho_f)]~=~\lambda_{\mathrm{max}}.
\end{equation}
Here, $\lambda_{\mathrm{max}}$ is the maximum element of the set $\{\lambda_n\}$. Hence, we infer that an initial state that overlaps with the final prepared state $|E_{n^{\ast}}\rangle$ maximally yields the minimal $T_{\mathrm{QSL}}$, which is both compelling and natural in retrospect. To determine the arrangement of other populations with $\lambda_{\mathrm{max}}$ excluded, we then turn to the dissipated heat $Q$,
\bea
Q &=& \mathrm{Tr}[H_S\rho_0]-E_{n^{\ast}}\nonumber\\
&=& \sum_{n\neq n^{\ast}}E_n\lambda_n+(\lambda_{\mathrm{max}}-1)E_{n^{\ast}}.
\eea
For a given set $\{\lambda_n\}$, the second term $(\lambda_{\mathrm{max}}-1)E_{n^{\ast}}$ on the right-hand-side of the above equation is fixed, therefore we only need to optimize the first term. Noting that the energy eigenvalues $\{E_n\}$ is sorted in an increasing order, according to the the Majorization theory (see, e.g., Ref. \cite{Mohammady.16.NJP}), the first term attains its minimum when the probabilities $\lambda_n$ are sorted in a decreasing order, $\lambda_1\geqslant\lambda_2\cdots\geqslant \lambda_N$. Putting together, we can conclude that an initial state with the optimal permutation of populations should have the following structure
\begin{equation}\label{eq:opt_rho}
\rho_0^{\mathrm{opt}}~=~\sum_{n\neq n^{\ast}}\lambda_n|E_n\rangle\langle E_n|+\lambda_{\mathrm{max}}|E_{n^{\ast}}\rangle\langle E_{n^{\ast}}|,
\end{equation}
where the first term on the right-hand-side resembles the structure of a passive state \cite{Allahverdyan.04.EPL,Uzdin.21.PRXQ} in the sense of the same orderings $\lambda_1\geqslant\lambda_2\cdots\geqslant \lambda_N$ and $E_1\leqslant E_2\cdots\leqslant E_N$; Noting $n^{\ast}$ is excluded. We remark that by writing down the above form we implicitly assume that the final prepared state is an excited state of $H_S$, namely, $n^{\ast}\neq 1$. We note that if the desired final state $|E_{n^{\ast}}\rangle$ happens to be the ground state of $H_S$, then the optimal initial state is strictly a passive one. 

\section{Example: Coupled Rydberg atom system}\label{sec:5}
To validate the above theoretical expectations, we consider a setup consisting of two $\Lambda$-type three-level Rydberg atoms, each contains two ground states $|0\rangle$ and $|1\rangle$, and one Rydberg state $|r\rangle$. It was recently showed \cite{LiD.18.PRA} that the setup can be utilized to realize the DSP of the Bell state $|\Phi\rangle=(|00\rangle-|11\rangle)/\sqrt{2}$ with $|00(11)\rangle$ being understood as $|0(1)\rangle\otimes|0(1)\rangle$ . The corresponding Lindblad master equation reads \cite{LiD.18.PRA}
\begin{equation}\label{eq:lindblad_R}
\dot{\rho}_t~=~-i[H_S,\rho_t]+\frac{\gamma}{2}\sum_{\mu=1}^4\left[L_{\mu}\rho_tL_{\mu}^{\dagger}-\frac{1}{2}\{L_{\mu}^{\dagger}L_{\mu},\rho_t\}\right].
\end{equation}
Here, the system Hamiltonian takes the following form 
\bea
H_S &=& \Omega_2(|10\rangle\langle r0|+|01\rangle\langle 0r|)+\omega\Big[(|11\rangle+|00\rangle)\nonumber\\
&&\otimes (\langle 01|+\langle 10|)\Big]+\mathrm{H.c.}
\eea
with $\mathrm{H.c.}$ denoting Hermitian conjugate, and we have four decaying channels with corresponding Lindblad jump operators
\bea
L_1 &=& |01\rangle\langle 0r|,~~L_2~=~|00\rangle\langle 0r|,\nonumber\\
L_3 &=& |10\rangle\langle r0|,~~L_4~=~|00\rangle\langle r0|.
\eea
$\gamma$ is a damping rate. One can easily check that the Bell state satisfies $H_S|\Phi\rangle=0$ and $L_{1,2,3,4}|\Phi\rangle=0$ as required by the DSP scheme. With the above Lindblad operators, we find $\mathcal{A}=\sqrt{2}\gamma/4$ in Eq. (\ref{eq:QSL}). In the following analysis, we adopt experimental values for the parameters $(\Omega_2,\omega,\gamma)=2\pi\times(0.02,0.01,0.03)$ MHz \cite{Grankin.14.NJP}. 

Under the adopted parameter set, we obtain via exact diagonalization the following ordered six energy eigenstates of $H_S$ expressed in the local basis of $\{|00\rangle,|01\rangle,|10\rangle,|11\rangle,|0r\rangle,|r0\rangle\}$,
\begin{flalign}
&|E_1\rangle =\frac{1}{\sqrt{8}}(|00\rangle+|11\rangle+|0r\rangle+|r0\rangle)-\frac{1}{2}(|01\rangle+|10\rangle),\nonumber\\
&|E_2\rangle=\frac{1}{2}(|01\rangle+|r0\rangle)-\frac{1}{2}(|10\rangle+|0r\rangle),\nonumber\\
&|E_3\rangle=-\frac{1}{2}(|00\rangle+|11\rangle)+\frac{1}{2}(|0r\rangle+|r0\rangle),\nonumber\\
&|E_4\rangle=\frac{1}{\sqrt{2}}(|00\rangle-|11\rangle)~=~|\Phi\rangle,\nonumber\\
&|E_5\rangle=\frac{1}{2}(|10\rangle+|r0\rangle)-\frac{1}{2}(|01\rangle+|0r\rangle),\nonumber\\
&|E_6\rangle=\frac{1}{\sqrt{8}}(|00\rangle+|11\rangle+|0r\rangle+|r0\rangle)+\frac{1}{2}(|01\rangle+|10\rangle).
\end{flalign}
We remark that the prepared Bell state corresponds to the fourth excited state of $H_S$. 

Since we are just interested in the permutations of a given set $\{\lambda_n\}$, we can take an arbitrary yet physical set of $\{\lambda_n\}$ for numerical demonstration. As an illustration, we first compare the following three permutations containing the same elements with unity sum with respect to the ordered energy basis $\{|E_n\rangle\}$: 
\bea\label{eq:abc}
\{\lambda_n^A\} &=& (0.2,0.15,0.1,0.4,0.08,0.07),\nonumber\\ 
\{\lambda_n^B\} &=& (0.07,0.08,0.1,0.15,0.2,0.4),\nonumber\\
\{\lambda_n^C\} &=& (0.4,0.2,0.15,0.1,0.08,0.07).
\eea
In this case, permutation $A$ emerges as the optimal choice among all permutations, as it exhibits the maximum population of $0.4$ on state $|E_4\rangle$, with other elements arranged in a decreasing order according to Eq. (\ref{eq:opt_rho}). On the other hand, permutation $C$ represents a passive state, while permutation $B$ is chosen due to its natural ordering, which serves as the inverse of its counterpart, $C$. We emphasize that there is nothing special about the chosen elements of $\{\lambda_n\}$ in Eq. (\ref{eq:abc}). We depict the evolution dynamics of the relative purity $\mathrm{Tr}[\rho_t\rho_f]$ in Fig. \ref{fig:2} obtained by solving the Lindblad master equation Eq. (\ref{eq:lindblad_R}) subject to initial states $\rho_0=\sum_n\lambda_n|E_n\rangle\langle E_n|$ generated by the permutations in Eq. (\ref{eq:abc}). We highlight the arrangements of Eq. (\ref{eq:abc}) on the top of Fig. \ref{fig:2}. In the inset we further list the values of the QSL $T_{\mathrm{QSL}}$ [cf. Eq. (\ref{eq:QSL})] and dissipated heat $Q$ [cf. Eq. (\ref{eq:heat})] for {\it all} possible permutations of the same elements as in Eq. (\ref{eq:abc}). 

As demonstrated in Fig. \ref{fig:2} and its inset, it becomes clear that the optimal permutation $A$ achieves the minimal QSL while also exhibiting the smallest dissipated heat among the set of permutations that yield the same QSL. This result illustrates the methodology behind our Pareto-type optimization strategy. In comparison, the permutation $\{\lambda_n^B\}$ fails to minimize both the QSL $T^{\mathrm{QSL}}$ and the dissipated heat $Q$. While the permutation $\{\lambda_n^C\}$ minimizes $Q$ as expected, it leads to the slowest evolution dynamics among the three scenarios in Eq. (\ref{eq:abc}). From the results, we remark that the dissipated heat $Q$ can be negative as the final prepared state is an excited eigenstate of the system, implying that the environment injects energy into the system to assist the DSP process. This seems at odds with the usual Landauer bound \cite{Landi.21.RMP} which states instead a positive $Q$ as $\Delta S$ is positive here. However, we should point out that the engineered bath used in DSP is usually non-thermal (see, e.g., Ref. \cite{Leghtas.15.S}) and one cannot assign a meaningful thermodynamic temperature to it. Without a temperature, it is difficult to relate heat to entropy production, hence the usual Landauer bound is not applicable. In this regard, a refined Landauer bound \cite{Timpanaro.20.PRL} may help.
%
\begin{figure}[thb!]
 \centering
\includegraphics[width=1\columnwidth]{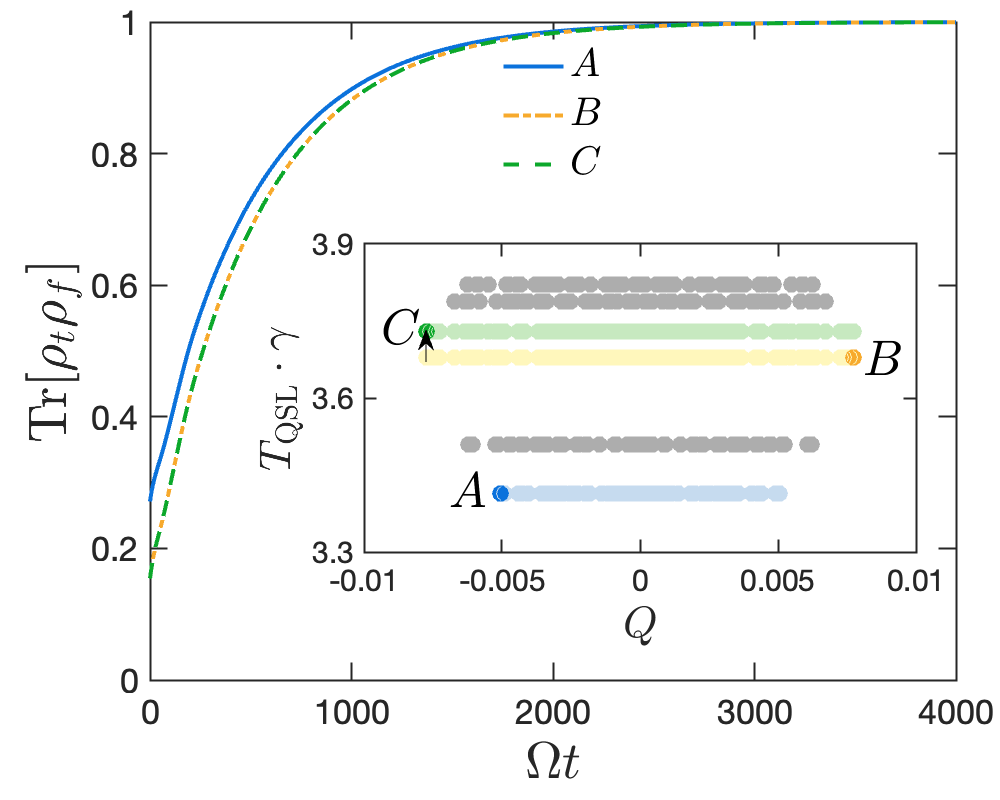}
\caption{Relative purity $\mathrm{Tr}[\rho_t\rho_f]$ as a function of evolution time obtained by solving Eq. (\ref{eq:lindblad_R}) with three initial conditions $\rho_0^{A,B,C}=\sum_n\lambda_{n,th}^{A,B,C}|E_n\rangle\langle E_n|$. Here, $\{\lambda_{n,th}^{A,B,C}\}$ are permutations of thermal populations $\{e^{-\beta E_n}/\sum_n{e^{-\beta E_n}}\}$ according to the same ordering rule in Eq. (\ref{eq:abc}). Permutation $A$ is again the optimal one. Inset: 
we show the corresponding values of $T_{\mathrm{QSL}}$ [cf. Eq. (\ref{eq:QSL})] and $Q$ [cf. Eq. (\ref{eq:heat})] for all possible permutations of the same set of thermal populations $\{e^{-\beta E_n}/\sum_n{e^{-\beta E_n}}\}$, with permutations $A,~B$ and $C$ highlighted. For a better view, we shift the data set containing permutation $C$ upwards by an amount of 0.05 $\gamma^{-1}$. In the calculation, we set $\Omega=2\pi$ MHz as the unit and take an inverse temperature $\beta=20$.
}
\protect\label{fig:3}
\end{figure}

Although we take an arbitrary set of populations in the above, we remark that the distinctions between three orderings $\{\lambda_n^{A,B,C}\}$ are independent of detailed values of $\{\lambda_n\}$. To see this, we instead consider populations $\{\lambda_{n,th}\}=\{e^{-\beta E_n}/\sum_n{e^{-\beta E_n}}\}$ generated from a thermal initial state $e^{-\beta H_S}/\mathrm{Tr}[e^{-\beta H_S}]$ with $\beta$ an inverse temperature; One can couple the system to a thermal bath initially to obtain such a thermal initial state. In Fig. \ref{fig:3}, we depict such a set of results obtained from three permutations $\{\lambda_{n,th}^{A,B,C}\}$ according to the same ordering rule in Eq. (\ref{eq:abc}) by fixing the temperature value. Compared with Fig. \ref{fig:2}, it is evident that one still observes the same behaviors of the dynamics of $\mathrm{Tr}[\rho_t\rho_f]$ as well as dissipated heat $Q$. Particularly, from the inset of Fig. \ref{fig:3}, we see that the permutation $A$ again minimizes the QSL together with a smallest $Q$ among the set with the same QSL. We note that for the chosen parameter set $E_3\simeq -1.05\times 10^{-18}$ and $E_4=0$ are very close in energy in our example, hence the permutations $\{\lambda_{n,th}^B\}$ and $\{\lambda_{n,th}^C\}$ sorted in respectively an increasing and a decreasing order in magnitude result in almost indistinguishable dynamical evolution of the relative purity $\mathrm{Tr}[\rho_t\rho_f]$ as can be seen from Fig. \ref{fig:3}. In the inset, we shift the data set containing the permutation $C$ upwards for a better view. Nevertheless, the dissipated heat $Q$ results from permutations $B$ and $C$ are different.

\section{Conclusion and outlook}\label{sec:6}
In this study, we derived an initial-state dependent quantum speed limit (QSL) for measuring the preparation time of dissipative state preparation (DSP) processes. Based on the obtained QSL, we performed a Pareto-type initial state optimization under a fixed set of populations in the system energy eigenbasis. By conditionally minimizing the dissipated heat based on an optimized QSL, we determined the optimal permutation of populations with respect to an ordered energy eigenbasis. Our findings revealed that the initial state corresponding to this optimal permutation exhibits the maximum overlap with the final prepared states among all possible states generated from permutations of populations. Furthermore, the remaining population elements are sorted in an order that resembles a passive state. We demonstrated the effectiveness of our framework in a model system consisting of two three-level Rydberg atoms used to dissipatively prepare the Bell state.

In the context of preparing desired states, it is commonly assumed that the process essentially involves cooling the system to the ground state of an appropriate many-body Hamiltonian. Hence the dissipated heat out of the system is always positive, just as existing qubit reset and information erasure studies \cite{Mohammady.16.NJP, Klaers.19.PRL, Miller.20.PRL, Zhen.21.PRL, Saito.22.PRL} showed. However, our numerical analysis of a dissipative Rydberg atom system for preparing the excited Bell state unveils a distinct thermodynamic behavior within the DSP scheme, namely, dissipated heat can be negative under optimal scenarios. This result suggests that the environment injects energy into the system to facilitate DSP processes, enabling rapid preparation of an excited state.

Furthermore, conventional expressions for total entropy production are not applicable in DSP \cite{Abe.03.PRA, Santos.17.PRL}, highlighting the need for a self-consistent quantum thermodynamic framework for DSP. The development of such a framework remains an area for future research and promises to be a valuable contribution to the field of dissipative quantum information processing. 

\begin{acknowledgments}
J. Liu acknowledges supports from Shanghai Pujiang Program (Grant No. 22PJ1403900), the National Natural Science Foundation of China (Grant No. 12205179), and start-up funding of Shanghai University. H. Nie is supported by CQT PhD programme.
\end{acknowledgments}

\appendix
\renewcommand{\theequation}{A\arabic{equation}}
\renewcommand{\thefigure}{A\arabic{figure}}
\setcounter{equation}{0}  
\setcounter{figure}{0}  
\section{Another derivation for quantum speed limit}
\label{a:1}
To obtain a QSL, one can directly start from the equation of motion for the fidelity \cite{Del.13.PRL},
\begin{equation}
\frac{d}{dt}\mathrm{Tr}[\rho_t\rho_f]~=~\mathrm{Tr}[\rho_f(\mathcal{L}\rho_t)]~=~\mathrm{Tr}[(\mathcal{L}^{\dagger}\rho_f)\rho_t]
\end{equation}
Here, we introduced Liouvillian superoperator $\mathcal{L}$ and its adjoint 
\begin{equation}
\mathcal{L}^{\dagger}\rho~=~i[H_S,\rho]+\sum_{\mu}\gamma_{\mu}\left[L_{\mu}^{\dagger}\rho L_{\mu}-\frac{1}{2}\{L_{\mu}^{\dagger}L_{\mu},\rho\}\right].
\end{equation}
Using properties of $\rho_f$, we find $\mathcal{L}^{\dagger}\rho_f=\sum_{\mu}\gamma_{\mu}L_{\mu}^{\dagger}\rho_fL_{\mu}$. Using the Cauchy-Schwarz inequality, we have
\begin{equation}
\left|\frac{d}{dt}\mathrm{Tr}[\rho_t\rho_f]\right|~\leqslant~\mathcal{A}||\rho_t||_F~\leqslant~\mathcal{A}.
\end{equation}
Recalled that $\mathcal{A}\equiv||\sum_{\mu}\gamma_{\mu}L_{\mu}^{\dagger}\rho_fL_{\mu}||_F$. Adopting the parametrization and integrating both sides of the above inequality, we receive
\begin{equation}
T~\geqslant~T_{\mathrm{QSL},2}~\equiv~\frac{\int_{\Theta_0}^{\Theta_f}d|\cos\Theta_t|}{\mathcal{A}}~=~\frac{1-\cos\Theta_0}{\mathcal{A}}.
\end{equation}
Since $\cos\Theta_0\in[0,1]$ with $\Theta_0\in[0,\pi/2]$, one can easily proof that $1-\cos\Theta_0\leqslant \sqrt{2(1-\cos\Theta_0)}$, therefore the bound $T_{\mathrm{QSL}}$ obtained in the main text is tighter than $T_{\mathrm{QSL},2}$ got here, namely,
\begin{equation}
T~\geqslant~T_{\mathrm{QSL}}~\geqslant~T_{\mathrm{QSL},2}.
\end{equation}

\renewcommand{\theequation}{B\arabic{equation}}
\renewcommand{\thefigure}{B\arabic{figure}}
\setcounter{equation}{0}  
\setcounter{figure}{0}  
\section{Passive initial states minimize the heat exchange alone}
\label{a:2}
Since the final state is fixed, we have
\begin{equation}
Q~\geqslant~\mathop{\mathrm{min}}\limits_{\rho_0}\left(\mathrm{Tr}[H_S\rho_0]\right)-E_{n^{\ast}}.
\end{equation}
To obtain the minimum of $\mathrm{Tr}[H_S\rho_0]$, we resort to the Majorization theory \cite{Mohammady.16.NJP}: Noting that the energy basis we considered is already ordered with ordered eigenvalues $E_1\leqslant E_2\leqslant\cdots\leqslant E_N$, the minimum of $\mathrm{Tr}[H_S\rho_0]$ is attained when we consider a passive initial state \cite{Allahverdyan.04.EPL,Uzdin.21.PRXQ}
\begin{equation}
\sigma_0~=~\sum_{n=1}^N\tilde{\lambda}_n|E_n\rangle\langle E_n|
\end{equation}
with $\{\tilde{\lambda}_n\}$ a permutation of $\{\lambda_n\}$ such that probabilities are sorted in a decreasing order, $\tilde{\lambda}_1\geqslant\tilde{\lambda}_2\geqslant\cdots\geqslant \tilde{\lambda}_N$. By doing so, we find
\bea
&&\mathop{\mathrm{min}}\limits_{\rho_0}\left(\mathrm{Tr}[H_S\rho_0]\right)~=~\mathrm{Tr}[H_S\sigma_0]~=~\sum_{n=1}^N\tilde{\lambda}_nE_n,\nonumber\\
&&\Longrightarrow Q~\geqslant~\sum_{n=1}^N\tilde{\lambda}_nE_n-E_{n^{\ast}}.
\eea
We remark that since the DQSP is a non-unitary process it is possible for a passive state to exchange heat with the environment. Now let us turn to the fidelity
\begin{equation}
\mathrm{Tr}[\sigma_0\rho_f]~=~\tilde{\lambda}_{n^{\ast}},
\end{equation}
which is in general smaller than $\tilde{\lambda}_1$ and hence $\sigma_0$ is not optimal for minimizing the QSL $T_{\mathrm{QSL}}$ among other permutations of $\{\lambda_n\}$.

%

\end{document}